\documentclass[sigconf, nonacm]{acmart}
\AtBeginDocument{%
  }

\usepackage{float}
\usepackage{amsmath}
\usepackage[utf8]{inputenc} 
\usepackage[T1]{fontenc}    
\usepackage{hyperref}       
\usepackage{url}            
\usepackage{booktabs}       
\usepackage{amsfonts}       
\usepackage{nicefrac}       
\usepackage{microtype}      
\usepackage{lipsum}
\usepackage{tabularray}
\usepackage{graphicx}       
\usepackage{ninecolors}
\usepackage{listings}
\usepackage{acronym}
\acrodef{RAG}[\texttt{RAG}]{Retrieval-Augmented Generation}
\acrodef{LLM}{Large Language Model}
\acrodef{QA}{Question-Answering}
\acrodef{RAGF}[\texttt{RAGF}]{\acs{RAG}-Fusion}
\acrodef{CRAG}{Corrective Retrieval Augmented Generation}
\acrodef{RRF}{recriprocal rank fusion}

\newcommand{\ragelo}{\texttt{RAGElo}}

\begin{document}
\title{Evaluating RAG-Fusion with \ragelo{}: an Automated Elo-based Framework}

\author{Zackary Rackauckas}
\affiliation{%
    \institution{Columbia University}
    \city{New York}
    \state{NY}
    \country{USA}
}
\authornote{Work conducted while the author was affiliated with Infineon Technologies}
\email{zcr2105@columbia.edu}

\author{Arthur C\^amara}
\affiliation{
    \institution{Zeta Alpha}
    \city{Amsterdam}
    \country{The Netherlands}
}
\email{camara@zeta-alpha.com}

\author{Jakub Zavrel}
\affiliation{
    \institution{Zeta Alpha}
    \city{Amsterdam}
    \country{The Netherlands}
}
\email{zavrel@zeta-alpha.com}

\begin{abstract}
Challenges in the automated evaluation of Retrieval-Augmented Generation (RAG) Question-answering (QA) systems include hallucination problems in domain-specific knowledge and the lack of gold standard benchmarks for company-internal tasks. This results in difficulties in evaluating RAG variations, like \ac{RAGF} in the context of a product QA task at Infineon Technologies. To solve these problems, we propose a comprehensive evaluation framework, which leverages \acfp{LLM} to generate large datasets of synthetic queries based on real user queries and in-domain documents, uses \emph{LLM-as-a-judge} to rate retrieved documents and answers, evaluates the quality of answers, and ranks different variants of \ac{RAG} agents with \ragelo{}'s automated Elo-based competition. \emph{LLM-as-a-judge} rating of a random sample of synthetic queries shows a moderate, positive correlation with domain expert scoring in relevance, accuracy, completeness, and precision. While \ac{RAGF} outperformed \ac{RAG} in Elo score, a significance analysis against expert annotations also shows that \ac{RAGF} significantly outperforms \ac{RAG} in completeness, but underperforms in precision. In addition, Infineon's RAGF assistant demonstrated slightly higher performance in document relevance based on MRR@5 scores. We find that \ragelo{} positively aligns with the preferences of human annotators, though due caution is still required. Finally, \ac{RAGF}'s approach leads to more complete answers based on expert annotations and better answers overall based on \ragelo{}'s evaluation criteria.
\end{abstract}

\maketitle


\section{Introduction}

The text-generating capabilities of \acp{LLM}, together with their text understanding abilities, have allowed conversational \ac{QA} systems to experience a considerable leap in performance, with near-human text quality and reasoning capabilities~\cite{brown2020Language}. However, these systems can be prone to hallucinations~\cite{xiao2021Hallucination,ji2023Survey}, as they sometimes produce seemingly plausible but factually incorrect answers.

The general inability of such models to identify unanswerable questions~\cite{yin2023large, amayuelas2023knowledgeofknowledge} can exacerbate hallucinations, especially in enterprise settings. In such scenarios, user questions may require specific domain knowledge to be answered properly. This knowledge is usually out-of-domain for most \acp{LLM}, but is present in private and confidential internal documents from the company. 

One such company is Infineon, a leading manufacturer of semiconductors. Given its wide range of equipment, information about its products is spread across multiple, highly technical documents, including datasheets and selection guides of hundreds of pages. Therefore, an internal retrieval augmented conversational \ac{QA} system was developed by Infineon for internal users such as account managers, field application engineers, and sales operations specialists. This system allows professionals to ask questions about products from the whole catalog while in the field.

One of the features of Infineon's conversational agent is the usage of \emph{\acf{RAGF}}, a technique for increasing the quality of the generated answers by generating variations of the user question and combining the rankings produced by these variations using rank-fusion methods (i.e., \ac{RRF}~\cite{cormack-2009-reciprocal}) into a ranking that has both more diverse and higher quality answers.

However, \emph{evaluating} these systems bring complications common to retrieval augmented agents, especially in enterprise settings, stemming from the lack of comprehensive test datasets. Ideally, such a test set would comprise a large set of real user questions from a query log, paired with ``golden answers'' provided by experts. The lack of such a test set leads to two main issues. First, evaluation of answers generated by \acp{LLM} by traditional n-gram evaluation metrics such as ROUGE~\cite{lin-2004-rouge}, BLEU~\cite{papineni-2002-bleu}, and METEOR~\cite{Lavie-2007-meteor} is not possible, given the lack of ground truth answers. Second, and as a consequence, evaluating the quality of the answers generated by the LLM systems would require in-domain experts (potentially from within the company) in a process that is both slow and costly~\cite{Ziying2018pairwisecrowd}.

One approach for tackling the lack of an extensive test set is to use synthetic queries generated by \acp{LLM} as a proxy of user queries~\cite{arabzadeh2024acomparisonof}. However, the lack of in-domain knowledge of \acp{LLM} makes queries naively generated by these models unreliable and prone to hallucinations, especially when generating queries about specific products and their specifications (c.f., Table~\ref{tab:base_queries} for examples of real user's questions submitted to the system). 

To solve this, we propose to use a process similar to InPars~\cite{jeronymo2023InParsv2} to create a set of synthetic evaluation queries. We ask \acp{LLM} to generate queries based on portions of existing documentation injected into the prompt. To increase similarity to real user queries, we include existing user questions as few-shot examples to the prompt. With this process, we are able to generate a large set of high-quality synthetic queries for evaluating our systems. Figure~\ref{fig:synt_query_gen} describes the process of generating synthetic queries and the output of a search agent. Table~\ref{tab:dev_queries} shows a sample of these queries.

To tackle the second issue, a lack of ground truth ``golden answers,'' we leverage an \emph{LLM-as-a-judge} process, where a strong \ac{LLM} is used to evaluate the quality of the answers generated by the \ac{RAG} agent's \ac{LLM}~\cite{huang2024anempiricalstudy}. We then follow the practice of judging generated answers in a \emph{pairwise} fashion~\cite{zheng2023judgingllmas}, prompting the judge \ac{LLM} to select the better answer between two candidates generated by different \ac{RAG} pipelines. (c.f.~Section 6 with details of our pipelines).

Finally, to mitigate the lack of in-domain knowledge of the judging \ac{LLM}, we also annotate the relevance of the documents retrieved by the pipelines being evaluated and inject the relevant documents in the context used by the judging \ac{LLM}. This allows the judging \ac{LLM} to better assess for hallucinations and completeness and better align the quality of the evaluations to those conducted by experts.

This process is mediated by \ragelo{}~\footnote{\url{https://github.com/zetaalphavector/ragelo}}, a toolkit for evaluating \ac{RAG} systems inspired by the Elo ranking system. \ragelo{} provides an easy-to-use CLI and Python library for using \acp{LLM} to evaluate retrieval results and answers produced by \ac{RAG} pipelines. By combining a retrieval evaluator, a pairwise answer annotator, and an Elo-inspired tournament, \ragelo{} leverages powerful \acp{LLM} to agnostically annotate \emph{and rank} different \ac{RAG} pipelines. We notice that, although noisy, the \ac{LLM} annotations generated by \ragelo{}~are generally well aligned with experts' judgments of relative system quality, allowing for fast experimentation and comparisons between different \ac{RAG} implementations without the frequent intervention of experts as annotators.

This paper evaluates multiple implementations of Infineon's retrieval augmented conversational agent using \ragelo{}: a traditional \acl{RAG} and a \acl{RAGF} implementation. \acl{RAGF} generates multiple variations of the user question and combines the rankings produced by these queries into a more diverse set of documents. The documents are then fed into the \ac{LLM}. We also analyze these same agents under a keyword-based retrieval regimen (i.e., the retriever uses BM25 to retrieve and rank documents), a dense retriever, and a hybrid retriever that combines the ranking generated by the BM25 and the dense retrievers using \ac{RRF}. Our goal is to answer the following questions:
\begin{itemize}
    \item Does the evaluation framework proposed by \ragelo{} align with the preferences of human annotators for answers generated by \ac{RAG}-based conversational agents?
    \item Does the \ac{RAGF} approach of submitting multiple variations of the user question and combining their rankings lead to better answers?
\end{itemize}

\begin{table}[ht]
\centering
 \caption{Sample of questions submitted by users to the Infineon RAG-Fusion system}\label{tab:base_queries}
  \begin{tblr}{
    width=\columnwidth,
    colspec={X[l, -1]},
    colsep=1pt,
    rowsep=0pt,
    row{odd}={bg=azure9},
    column{1} = {cmd=\small},
    row{1} = {c, bg=azure3, fg=white, font=\sffamily, cmd=\textbf},
    hline{1,Z} = {solid, 1pt},
  }
    User-submitted queries\\
    What is the country of origin of IM72D128, and how does geopolitical exposure affect the market and my SAM for the microphone?\\
    What is the IP rating of mounted IM72D128?\\
    Tell me microphones that have been released since January 2023 based on the datasheet revision history.\\
    We need to confirm whether the IFX waterproof MIC has a sleeping mode and wake-up functions.
  \end{tblr}%
\end{table}
\begin{table}[ht]
\centering
 \caption{Sample of synthetic queries for evaluating Infineon's \ac{RAG} assistant. GPT4 refers to OpenAI's \texttt{\small gpt-4-turbo-2024-04-09} model. Opus, Sonnet and Haiku refer to Anthropic's Claude 3 models \texttt{\small opus-20240229}, \texttt{\small sonnet-20240229} and \texttt{\small haiku-20240307}, respectively.}\label{tab:dev_queries}
  \begin{tblr}{
    width=\columnwidth,
    colspec={X[l, -1] X[l, -1]},
    colsep=1pt,
    rowsep=0pt,
    row{odd}={bg=azure9},
    column{1} = {cmd=\small},
    row{1} = {c, bg=azure3, fg=white, font=\sffamily, cmd=\textbf},
    hline{1,Z} = {solid, 1pt},
  }
    model   & Query \\
    GPT4 & What are some typical consumer applications for TLV496x-xTA/B sensors?\\
    GPT4 & What specific ISO 26262 readiness is available for the KP253 sensor?\\
    Opus & How small of a form factor can I achieve for a battery-powered air quality device using Infineon's PAS CO2 sensor?\\
    Sonnet & Can Infineon's sensors support bus configurations or daisy-chaining for simplified wiring and reduced complexity in IoT systems?\\
    Haiku & Which TLE4971 current sensor models are available in the TISON-8-6 package?\\
  \end{tblr}%
\end{table}

\section{Related Work}
\begin{figure*}[ht!]
\centering
\includegraphics[width=.95\linewidth]{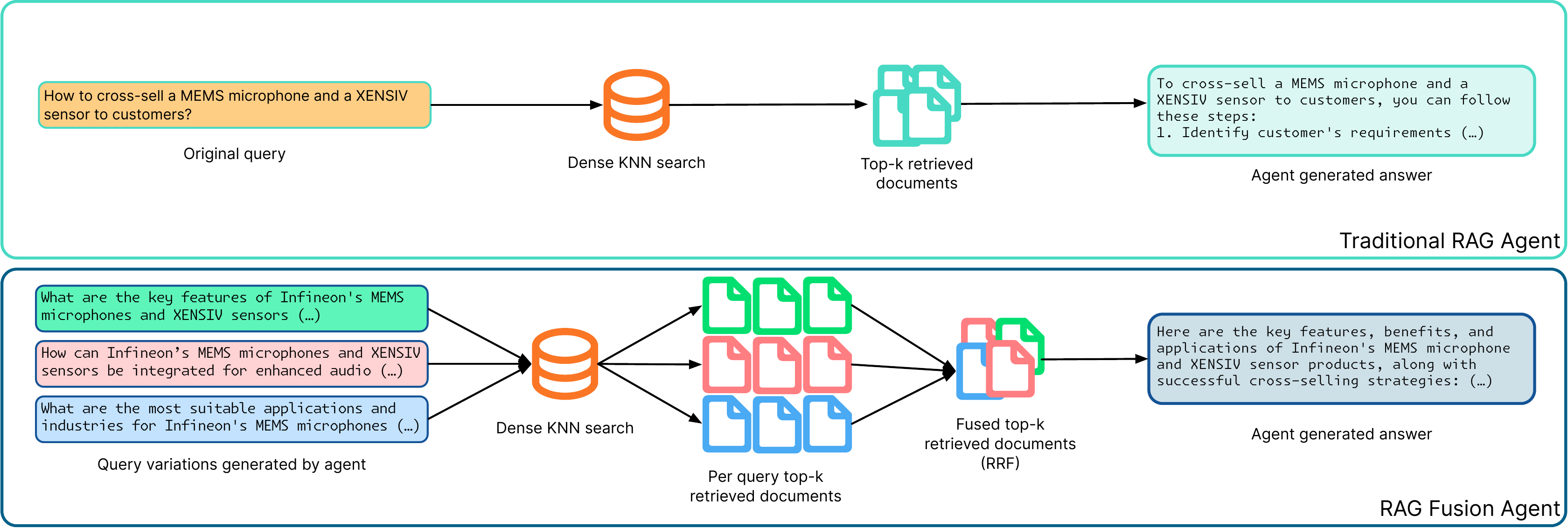}
\caption{A traditional \acl{RAG} pipeline compared to a \acl{RAGF} pipeline. While a traditional \ac{RAG} agent submits only the original query to the search system, a \ac{RAGF} agent first generates variations of the user query and combines the rankings induced by these queries into a final ranking using RRF. The resulting top-k passages are fed into the \ac{LLM} for generating the answer to the user's query.}\label{fig:agents}
\end{figure*}

Several evaluation systems for \ac{RAG} have been proposed to address flaws in current evaluation methods. For instance, Facts as a Function (FaaF)~\cite{katranidis2024faaf} is an end-to-end factual evaluation algorithm specially created for \ac{RAG} pipelines. By creating functions from ground truth facts, FaaF focuses on the quality of generation and retrieval by calling \acp{LLM}. FaaF has substantially increased efficiency and cost-effectiveness, achieving reduced error rates compared to traditional evaluation methods. The reliance on a set of ground truths does not meet our goal of applying an automated evaluation toolkit to our pipelines. Recently, researchers have moved to eliminate the need for ground truths. This is especially important when automatically evaluating agents that retrieve highly technical documents from a large database, such as the Infineon \ac{RAGF} conversational agent. \ragelo{} eliminates this reliance by using an \emph{LLM-as-a-judge}, a method studied in numerous recent works.

SelfCheckGPT demonstrates the ability to leverage \acp{LLM} to detect and rank factuality with zero resources \cite{manakul2023selfcheckgpt}. In addition, it has been demonstrated that GPT3.5 Turbo outperforms ground truth baselines in fact-checking with a "1/2-shot" method \cite{zhang2023interpretable}. A model built to classify statements as true or false based on the activations of an \ac{LLM}'s hidden layers had up to 83$\%$ classification accuracy \cite{azaria2023internal}. This evidence supports \ragelo{}'s usage of \emph{LLM-as-a-judge}.

Automated evaluation metrics can also be applied to RAG-based agents. BARTScore, an automated metric based on the BART architecture, has also outperformed most metrics on categories including factuality \cite{lewis2019bart, NEURIPS2021_e4d2b6e6}. Besides automated evaluation metrics, several automated evaluation frameworks have been created with a similar goal to \ragelo{}. Focusing on faithfulness, answer relevance, and content relevance, RAGAS leverages LLM prompting to focus on situations where ground truths and human annotations are not present in a dataset \cite{es2023ragas}. Prediction-powered inference aims to decrease the number of human annotations needed for machine learning prediction on a dataset of images of galaxies with approximately 300,000 annotations \cite{angelopoulos2023predictionpowered}. The ARES toolkit leverages prediction-powered inference to evaluate \ac{RAG} systems with fewer human annotations. Like~\ragelo{}, ARES automatically evaluates \ac{RAG} systems using synthetically generated data \cite{saadfalcon2024ares}.

ARAGOG highlights Hypothetical Document Embedding (HyDE) and LLM reranking as effective methods for enhancing retrieval precision while also exploring the effectiveness of Sentence Window Retrieval and the potential of the Document Summary Index in improving \ac{RAG} systems \cite{eibich2024aragog}.

While the aforementioned frameworks evaluate answers on relevance, faithfulness, and correctness metrics, RAG can also be evaluated on noise and counterfactual robustness, negative rejection, and information integration \cite{chen2023benchmarking}.

In addition to answers, frameworks have also been created to evaluate documents. Corrective Retrieval Augmented Generation (CRAG) builds on \ac{RAG} by employing a retrieval evaluator to ensure that only the optimal documents are fed into the \ac{LLM} prompt prior to the answer generation phase~\cite{yan2024corrective}.

Due to its Elo-based ranking system for answers, its use of \emph{LLM-as-a-judge}, and its relevance evaluation of the intermediate retrieval steps in a \ac{RAG} pipeline, \ragelo{} is a unique evaluation toolkit. In this study, we use it to compare a simple \ac{RAG} versus a more sophisticated \ac{RAGF} system on a knowledge-intensive industry-specific domain.

\section{Retrieval Augmented QA with rank fusion}
While answers generated by traditional retrieval augmented systems are based on a number of documents retrieved from a single query, \ac{RAGF} introduces additional variation into the retrieval process. Upon receiving a query from the user, a \ac{RAGF} agent leverages a large language model to generate a set of queries based on the original \cite{Fazlija2024thesis}. Table~\ref{tab:fusion_queries} shows examples of queries generated by the agent based on the query, ``How to cross-sell a MEMS microphone and a XENSIV sensor to customers?''.

\begin{table}[ht]
\centering
 \caption{Queries Generated from ``How to cross-sell a MEMS microphone and a XENSIV sensor to customers?''}\label{tab:fusion_queries}
  \begin{tblr}{
    width=\columnwidth,
    colspec={X[l, -1]},
    colsep=1pt,
    rowsep=0pt,
    row{odd}={bg=azure9},
    column{1} = {cmd=\small},
    row{1} = {c, bg=azure3, fg=white, font=\sffamily, cmd=\textbf},
    hline{1,Z} = {solid, 1pt},
  }
    LLM-Generated Query\\
    What are the key features of Infineon's MEMS microphones and XENSIV sensors that can be highlighted while cross-selling? \\
    How can Infineon's MEMS microphones and XENSIV sensors be integrated for enhanced audio and motion sensing capabilities in various applications? \\
    What are the most suitable applications and industries for Infineon's MEMS microphones and XENSIV sensors to maximize cross-selling potential? \\
  \end{tblr}%
\end{table}
 
After generating the variations for the user query, the \ac{RAGF} agent submits the original and the generated queries to a retrieval system~\cite{RAGIFX} that returns the top-$k$ relevant documents $d, d_1, d_2, \ldots d_k$ from the set of all documents $D$ for each query. The rankings induced by these queries are then combined using \acf{RRF}~\cite{cormack-2009-reciprocal} into a final, higher-quality set of passages. The intuition behind \ac{RAGF} is that submitting variations of the same query and combining the final rankings increases the likelihood of relevant passages being injected into the \ac{LLM} prompt. In contrast, non-relevant passages retrieved by a single query are discarded \cite{author_last_name2024ragfusion}. Figure~\ref{fig:agents} describes how \ac{RAG} and \ac{RAGF} differ.
\begin{equation} \label{eq:rrf}
    RRFScore(d \in D) = \sum_{r \in R}{\frac{1}{r(d) + k}}. 
\end{equation}

\section{Development of a synthetic test set}
\begin{figure}[ht]
\centering
\includegraphics[width=.45\textwidth]{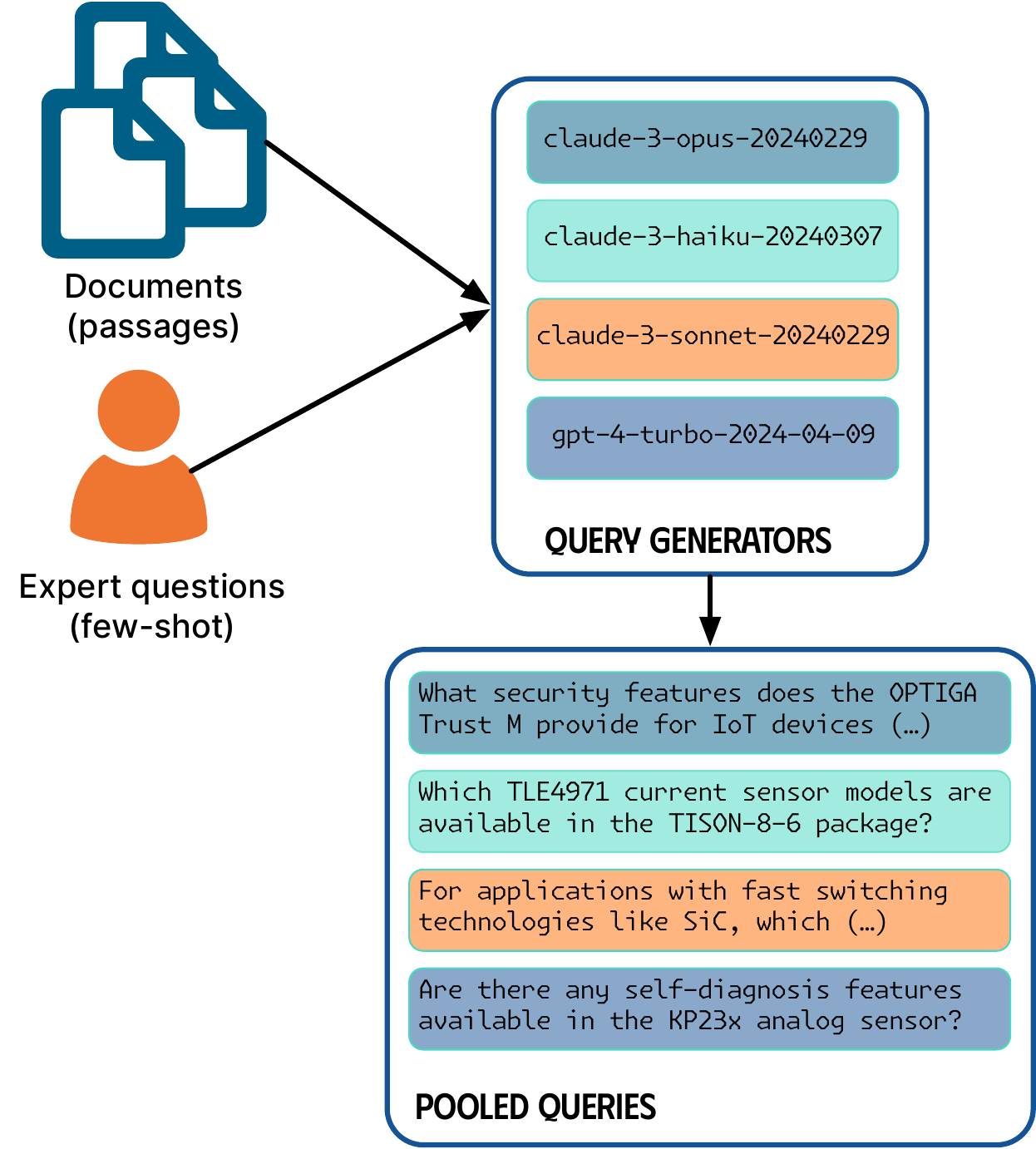}
\caption{Process for creating synthetic queries. We prompt multiple \acp{LLM} to generate queries based on existing documents. We include some existing user queries in the prompt as few-shot examples.}\label{fig:synt_query_gen}
\end{figure}

As previously discussed, one of the main issues when evaluating the quality of a \ac{QA} system in an enterprise setting is that, frequently, companies do not have a large enough existing collection of queries to evaluate such systems' quality. Therefore, in this work, we propose to adopt a strategy previously used by methods for generating synthetic queries for training retrieval systems, such as InPars~\cite{jeronymo2023InParsv2} and Promptagator~\cite{dai2022promptagator}. 

Similar to these approaches, we randomly sample passages from documents within our collection and prompt an \ac{LLM} to generate questions that users may ask about these portions. However, one difference in our approach to generating training queries is the size of these passages. When generating queries for training a retrieval system, we ideally want to keep the passages short to fit in the dense encoder's relatively short context windows. However, when generating queries for evaluating \ac{QA} systems (including retrieval augmented), we are not bound to the limit of the embedding model used for retrieval. Rather, a longer passage may yield questions that require multiple shorter passages to be answered. Therefore, we submit relatively long passages to the \acp{LLM}. Specifically, each passage is extracted from up to ten pages of PDF documents (about 2000 tokens~\footnote{All \acp{LLM} used in our experiments had long context windows of 128k or 200k tokens.})

To keep the questions generated as diverse as possible, we prompt four different \acp{LLM} to generate up to ten questions based on the same documents. Our test set collection contains a mix of queries generated by OpenAI's GPT-4 turbo \cite{gpt4turbo} and Anthropic's Claude-3 \cite{claude2024} Opus, Sonnet, and Haiku models~\footnote{We did not use GPT-3.5 or open source models due to their shorter context window at the time of writing.}. From a set of $N=840$ queries, we sampled $200$ queries across all four models. Half of the queries are selected from GPT-4 generated queries, and the other half from Claude 3 queries. Among the Claude 3 queries, to ensure the quality of the queries and their diversity, we again sample according to each model size. Ultimately, our test set contains 100 queries from GPT-4-turbo, 50 from Claude 3 Opus, 30 from Sonnet, and 20 from Haiku. 

Finally, to increase the quality of the generated queries, We asked for an account manager, a sales operations specialist, a marketing representative, and a business development manager to create queries that they would submit to the conversational agent from the perspective of their role. They were instructed to produce queries regarding products from the XENSIV sensor product line, consisting of MEM microphones, radar, current, magnetic, pressure, and environmental sensors. We compiled a list of 23 of these queries to use as a base for experimentation and used them as few-shot examples in the query generation prompt. Figure~\ref{fig:synt_query_gen} illustrates our method for generating synthetic queries based on existing user queries and document passages.

\section{LLM-as-a-Judge for \ac{RAG} pipelines}
\begin{figure*}[ht!]
\centering
\includegraphics[width=.95\linewidth]{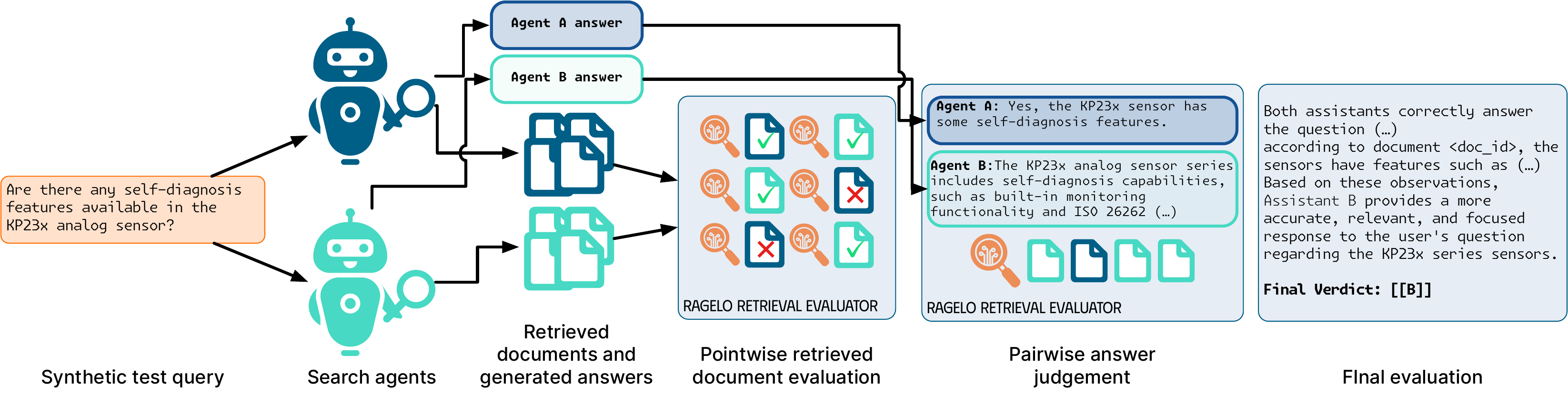}
\caption{The \ragelo{} evaluation pipeline. First, documents retrieved by the agents are evaluated pointwise according to their relevance to the user's question. Then, the agents' answers are evaluated pairwise, using the retrieved relevant documents from both agents as reference.}\label{fig:ragelo}
\end{figure*}
Even with a suitable set of synthetic questions for evaluating our \ac{RAG} conversational agent, assessing whether a given answer properly answers a question is not trivially done. If a ground-truth ``golden answer'' is available, one can use traditional syntactic-based metrics such as BLEU, METEOR or ROUGE~\cite{papineni-2002-bleu,Lavie-2007-meteor,lin-2004-rouge}. Without such reference answers, one would require human annotators with a considerable understanding of the question's topic to manually assess the quality of the answers produced by each system. However, this is a costly process. 

Alternatively, several LLM-as-a-Judge methods have been proposed, where another \ac{LLM} is asked to evaluate the quality of answers generated by other \acp{LLM}. Nevertheless, in an enterprise setting, the answers usually require the \ac{LLM} to access knowledge not present in their training datasets but rather contained in documents internal to the company. This is usually accomplished using a \ac{RAG} pipeline like the one described above. Therefore, the \emph{judging} \ac{LLM} also needs access to similar knowledge to accurately evaluate the agent's answers' quality.

Therefore, in this work, we rely on \ragelo{}, an open-source \ac{RAG} evaluation toolkit that evaluates the answers generated by each agent and the documents retrieved by them. By injecting the annotation of retrieved documents, pooled by the agents being evaluated, on the answer evaluation step, this method allows for the judging \ac{LLM} to evaluate if the generated answer was able to use all the information available about the question properly and to check for any hallucinations. As the documents used for generating the answers are included in the answer evaluating prompt, an agent that incorrectly cites information from a source or refers to information not present in these documents is likely hallucinating and should have its evaluation adjusted accordingly. As we explore in Section~\ref{sec:results}, this two-step process results in a high correlation between human expert annotators and the judging \ac{LLM}, enabling higher reliability and trust when evaluating different \ac{RAG} pipelines. This process is also illustrated in Figure~\ref{fig:ragelo}.

\subsection{Evaluation aspects}\label{sec:eval_aspects}
While our main evaluation focuses on the pairwise comparison between the two agents, \ragelo{} also allows us to evaluate answers pointwise. In this setting, similar to other works~\cite{thomas2023large}, we prompt the judging \ac{LLM} to evaluate the answers according to multiple criteria:
\begin{itemize}
    \item \textbf{Relevance:} Does the answer address the user's question?
    \item \textbf{Accuracy:} Is the answer factually correct, based on the documents provided?
    \item \textbf{Completeness:} Does the answer provide all the information needed to answer the user's question?
    \item \textbf{Precision:} If the user's question is about a specific product, does the answer provide the answer for that specific product?
\end{itemize}

\section{Retrieval pipelines}
We not only experiment with different search agents (i.e., \ac{RAG} and \ac{RAGF}. We are also interested in how different \emph{retrieval methods} may impact the quality of the final answers generated by these agents.

\subsection{Retrieval methods} Our corpus consists of passages extracted from the Infineon XENSIV Product Selection Guide, a 117-page document with detailed information on every product in the XENSIV family. This document included technical information about all Infineon XENSIV sensors, consumer and automotive sensor applications, guidance in selecting the correct sensor, and other comprehensive and detailed information about the product line. 

The passages are embedded using \textit{multilingual-e5-base}~ \cite{wang2024multilingual}~\footnote{\url{https://huggingface.co/intfloat/multilingual-e5-base}} and indexed using OpenSearch, allowing us to perform both KNN-based vector search, keyword-based search with BM25~\cite{robertson-1994-okapi}, and RRF based hybrids thereof.


\subsection{QA Systems Implementation} We mainly evaluate two agents: a naive \ac{RAG} pipeline, where the agent first retrieves top-$k$ passages that are then templated into a prompt, and the Infineon \acf{RAGF} agent. Upon receiving a query, a naive \ac{RAG} agent takes the following actions:

\begin{enumerate}
\item Retrieve the top \textit{k} most relevant passages from the search system.
\item Perform a Chat Completions API call, prompting the \ac{LLM} with instructions for generating an answer based on the five relevant passages.
\item Process and output the Chat Completions response.
\end{enumerate}

Meanwhile, the Infineon \ac{RAGF} conversational assistant uses a similar framework and performs the following steps upon receiving a query:
\begin{enumerate}
\item Perform a Chat Completions API call to generate four new queries based on the original query using a prompt tailored to the agent's original goal.
\item Retrieve the top \textit{k} most relevant passages for each query.
\item Using \ac{RRF}, combines the top-$k$ passages induced by all queries into a final ranking.
\item Perform a Chat Completions API call prompting the \ac{LLM} with carefully worded instructions for generating an answer based on the top-$k$ fused passages
\item Process and output the Chat Completions response.
\end{enumerate}


\section{Experiments}
\subsection{Comparing LLM-as-a-judge to expert annotators}
While \emph{LLM-as-a-judge} is a theoretically viable algorithm for rating \ac{RAG} and \ac{RAGF} answers, we must establish whether the results agree with the annotations of domain experts. 

\begin{figure}[ht]
\centering
\includegraphics[width=.45\textwidth]{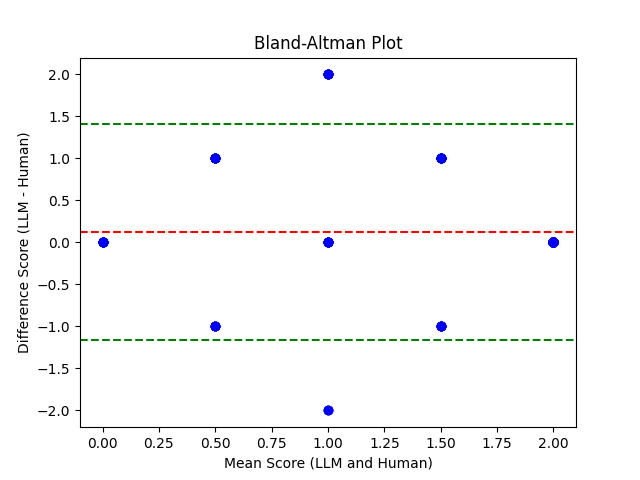}
\caption{Bland-Altman plot to visualize the comparison between \emph{LLM-as-a-judge} and expert answers.}\label{fig:bland-altman-fig}
\end{figure}

Figure \ref{fig:bland-altman-fig} provides a Bland-Altman plot to visually represent the LLM and human judgments' agreement. 

The bias of approximately $0.12$ indicates that, on average, \ac{LLM} scores were slightly higher than human scores. The limits of agreement ranged from approximately -1.17 to 1.41. demonstrating substantial variability in the difference between LLM and human evaluators.

Next, we compared \emph{LLM-as-a-judge} to expert annotators with Kendall's $\tau$. Kendall’s $\tau$ is a nonparametric measure that quantifies the degree of association between two monotonic continuous or ordinal variables by calculating the proportion of concordance and discordance among pairwise ranks, offering valuable insight into their rank correlation \cite{edwards2023graphing, perreault2022efficient}. We used the SciPy Stats Kendalltau function to calculate a tau-b score and a p-value for the combined ratings of all columns, flattened into a 1-D array with \ac{RAG} and \ac{RAGF} ratings combined \cite{scipy_kendalltau}. The tau-b value, a nonparametric measure of association, is calculated using the following formula \cite{kendall1945the}:
\begin{equation}
\tau_{b} = \frac{(P - Q)}{\sqrt{(P + Q + T) \cdot (P + Q + U)}}
\end{equation}
P represents the number of concordant pairs, Q represents the number of discordant pairs, T represents the number of ties exclusive to x, and U represents the number of ties exclusive to y.

This test returned $\tau \approx 0.56,$ indicating a moderate, positive correlation \cite{patrick2018correlation} with a p-value against a null hypothesis of no association of $p < 0.01$ ($99.99\%$ confidence level). For comparison, in similar experiments judging human versus LLM judgments, Faggioli et al. found $\tau$ values of $\tau = 0.76$ and $\tau = 0.86$\cite{Faggioli_2023}.

Following the same methodology, we also calculated Spearman's $\rho$, a similar nonparametric correlation measure. This resulted in $\rho \approx 0.59$ with $p < 0.01,$ demonstrating a statistically significant, moderate positive correlation \cite{perreault2022efficient}.

\subsection{\ac{RAG} vs \ac{RAGF}}
\subsubsection{Quality of retrieved documents}
We assessed document retrieval quality using Mean Reciprocal Rank@5 (MRR@5), which averages the inverse ranks of the first relevant result within the top five positions across all queries. The formula is given by 
\begin{equation}
MRR\mbox{@}5 = \frac{1}{|Q|} \sum_{i=1}^{|Q|}\frac{1}{\mbox{rank}_i},
\end{equation}
where $|Q|$ is the total number of queries and $\mbox{rank}_i$ is considered only if it's within the top five, otherwise it counts as zero \cite{jadon2024comprehensive}.

MRR@5 scores were calculated for each agent and each retrieval method considering two categories:
\begin{enumerate}
\item MRR@5 score for documents deemed ``somewhat relevant'' or ``very relevant.''
\item MRR@5 score for documents deemed ``very relevant.''
\end{enumerate}

The results can be seen below in Table~\ref{tab:mrr5_scores}. 

\begin{table}[ht]
\centering
 \caption{Mean MRR@5 scores for RAG vs RAG-F. The retrieval method columns indicate if the retrieval component used was vector search only (KNN), keywords only (BM25) or hybrid (KNN and BM25, combined with RRF).}\label{tab:mrr5_scores}
  \begin{tblr}{
    width=\columnwidth,
    colspec={X[l, c] X[l, c] X[l,c] X[l,c]},
    colsep=1pt,
    rowsep=0pt,
    row{odd}={bg=azure9},
    column{1} = {cmd=\small},
    row{1} = {c, bg=azure3, fg=white, font=\sffamily, cmd=\textbf},
    hline{1,Z} = {solid, 1pt},
  }
    Agent & Retrieval Method & Very Relevant & Somewhat Relevant \\
    RAG   & KNN     &  0.407 &  0.828 \\
    RAG   & BM25    &  0.821 &  0.955 \\
    RAG   & Hybrid  &  0.746 &  0.949 \\
    RAG-F & KNN     &  0.396 &  0.810 \\
    RAG-F & BM25    &  0.855 &  0.970 \\
    RAG-F & Hybrid  &  0.758 &  0.961 \\
  \end{tblr}%
\end{table} 

\subsubsection{Pairwise evaluation of answers}\label{sec:pairwise_eval}

We then ran \ragelo{} games to evaluate end-to-end answer quality of~\ac{RAG} vs~\ac{RAGF} with different base retriever configurations a task that cannot rely on standard Information Retrieval metrics. These \ragelo{} results show more victories for \ac{RAGF} than \ac{RAG}; For example, when using BM25 as a base retriever, \ac{RAG}F won 49\% of the games,~\ac{RAG} won 14.5\%, and \ac{RAG} and \ac{RAGF} are tied in 36.5\% of the times. The resulting Elo scores for all six variants are shown in table~\ref{tab:Elo_scores}, which give a robust ranking of the systems, without reliance on a gold standard. It is interesting to see, that for both \ac{RAGF} as well as \ac{RAG}, BM25 is a strong baseline that is not surpassed by generic embeddings in these experiments. 

Next, we compared the \ragelo{} outcome to the preference of our Infineon human annotator. We performed two-tailed paired t-tests to compare \ac{RAG} against \ac{RAGF} on each category from the Infineon representatives' human evaluations with $\alpha = .05$. As expected, due to its larger variety of retrieved results, \ac{RAGF} significantly outperforms \ac{RAG} in completeness at the 95\% confidence level with $p\approx0.01$ However, on the precision of answers, \ac{RAG} significantly outperformed \ac{RAGF} at the 95\% confidence level with $p\approx0.04.$

\begin{table*}[ht]
\centering
 \caption{RAG vs \ac{RAGF} Win percentage between pairwise comparison of the agent's answers using GPT-4o as a judge with \ragelo{}.}\label{tab:ragelo_results}
  \begin{tblr}{
    colspec={X[c] X[c] X[c] X[c] X[c] X[c] X[c] X[c] X[c]},
    width=\textwidth,
    colsep=1pt,
    rowsep=0pt,
    row{odd}={bg=azure9},
    row{1,2} = {c, bg=azure3, fg=white, font=\sffamily, cmd=\textbf},
    column{1,2} = {c, bg=azure3, fg=white, font=\sffamily, cmd=\textbf},
    hline{1,Z} = {solid, 1pt},
    hline{5,7,9} = {3-Z}{solid, 0.5pt},
    cell{1}{1} = {r=2,c=2}{c},
    cell{1}{3,5,7} = {r=1,c=2}{c},
    cell{3,5,7}{1} = {r=2,c=1}{c},
    cell{1}{9} = {r=2,c=1}{c},
    column{9} = {cmd=\textbf},
  }
  Agent     &           & BM25      &           & KNN       &           & Hybrid    &           & AVG    \\
            &           & RAG       & \ac{RAGF}     & RAG       & \ac{RAGF}     & RAG       & \ac{RAGF}     & \\
  BM25      & RAG       & ---       & 14.5\%    & 49.5\%    & 52.5\%    & 29.0\%    & 28.5\%    & 34.8\% \\
            & \ac{RAGF}     & 49.0\%    & ---       & 58.5\%    & 51.5\%    & 53.5\%    & 30.5\%    & 48.6\% \\
  KNN       & RAG       & 33.0\%    & 27.0\%    & ---       & 20.0\%    & 26.0\%    & 31.0\%    & 27.4\%  \\
            & \ac{RAGF}     & 34.5\%    & 30.0\%    & 37.0\%    & ---       & 30.5\%    & 32.0\%    & 32.8\%  \\
  Hybrid    & RAG       & 41.5\%    & 21.0\%    & 51.5\%    & 48.0\%    & ---       & 20.5\%    & 36.0\% \\
            & \ac{RAGF}     & 46.0\%    & 35.0\%    & 49.0\%    & 45.5\%    & 43.5\%    & ---       & 44.3\% \\     
  \end{tblr}%
    
\end{table*}

\begin{table}[ht]
\centering
 \caption{Elo Ranking for all agents averaged over 500 tournaments.}\label{tab:Elo_scores}
  \begin{tblr}{
  colspec={X[l] X[l] X[c]},
  width=\columnwidth,
    colsep=1pt,
    rowsep=0pt,
    row{odd}={bg=azure9},
    row{1} = {c, bg=azure3, fg=white, font=\sffamily, cmd=\textbf},
    column{1,2} = {font=\sffamily},
    hline{1,Z} = {solid, 1pt},
  }
Agent & Retrieval & Elo score  \\
\ac{RAGF}   & BM25      & 571.0 \\
\ac{RAGF}   & Hybrid    & 550.0 \\
RAG     & Hybrid    & 497.0 \\
RAG     & BM25      & 487.0 \\
\ac{RAGF}   & KNN       & 470.0 \\
RAG     & KNN       & 436.0 \\
\end{tblr}%
\end{table}



\section{Discussion}\label{sec:results}
As observed above, we found statistically significant, moderate positive correlations between LLM ratings and human annotations. This indicates a consistent association between the ratings from \emph{LLM-as-a-judge} and those by Infineon experts.
We find that on average, LLM scores are slightly higher than those of human annotators. This means that while relevance judgements on individual queries should not be fully reliable, and IR metrics derived from \emph{LLM-as-a-judge} should not be equated with regular relevance scores without further calibration, we can still make good use of this approach to rank-order systems. These findings collectively support the validity of our LLM evaluation method, which assesses conversational system outputs based on a combination of relevance, accuracy, completeness, and recall. 




The style of evaluation and the different dimensions it takes into account are specified in the prompts given to the LLM in the \ragelo{} evaluation, which are provided in Appendix~\ref{sec:prompts}. Specifically, while the initial \emph{LLM-as-a-judge} is given specific criteria to focus on only four categories, we instructed \ragelo{}'s impartial judge LLM to value more than the initial four categories: 
\begin{quote}
\textit{Your evaluation should consider factors such as comprehensiveness, correctness, helpfulness, completeness, accuracy, depth, and level of detail of their responses.}
\end{quote}

Since \ac{RAGF} significantly outperformed \ac{RAG} in the completeness category, the \ragelo{} judge LLM likely weighed completeness higher than precision. In addition, based on manual observation of a small random sample of answers, \ac{RAGF} produced more comprehensive answers and featured higher depth and level of detail due to the multiple query generation. However, games where \ac{RAG} won were most likely influenced by a significantly more precise answer than that of \ac{RAGF}.
While \ac{RAGF} values comprehensive answers that offer multiple perspectives to the user, \ac{RAG} produces shorter answers that answer the original query only. Since completeness is defined as the extent to which a user's question was answered, it can be presumed that \ac{RAGF}'s longer and more comprehensive answers may tend to be more complete. And since precision relates to the agent mentioning the correct product or product family, it can be presumed that \ac{RAGF}'s longer answers have more room to consider other products or product families, leading to reduced answer precision. While the human annotation was done by Infineon experts, different humans may rate answers differently, even if following the same set of criteria. 

A larger number of documents or a database of non-technical documents may have led to a different outcome. \ac{RAGF} can be applied to not only Infineon documents but also any documents database to retrieve. This includes not only enterprise uses but also uses in education, such as mathematics and language learning. The algorithm can be tuned to different use cases by tweaking the internal LLM prompt. For example, the Infineon \ac{RAGF} bot was prompted to "think like an engineer." However, an educator \ac{RAGF} bot could be prompted to "think like a teacher." Future work includes exploring other applications of \ac{RAGF}, especially in education. In addition, we will experiment with different prompts for both \emph{LLM-as-a-judge} and \ragelo{} while using different quantities and types of documents with the same retrieval algorithms.

Based on the calculated MMR@5 scores, we found that the \ac{RAGF} agent mostly outperforms the \ac{RAG} agent in ranking both highly relevant and somewhat relevant documents retrieved. This evidence search on multiple query variants produced, on average, slightly more higher-ranked relevant documents than using only the original user query. We also see that using vector search with embeddings is not a silver bullet, as for our test queries, BM25 seriously outperforms it. Since retrieval quality is highly dependent on the quality of the embeddings and their fit to the domain, this outcome will likely be changed by fine-tuning the embeddings, and adding additional intelligent re-rankers, which we leave here for future work, as the evaluation framework would remain the same.

\section{Conclusion}

Overall, we found that the evaluation framework proposed by \ragelo{} positively aligns with the preferences of human annotators for RAG and \ac{RAGF} with due caution due to a moderate correlation and variability of scoring. We found that the \ac{RAGF} approach leads to better answers most of the time, according to the \ragelo{} evaluation. According to expert scoring, the \ac{RAGF} approach significantly outperforms in completeness compared to RAG but significantly underperforms in precision compared to RAG. Based on these results, we cannot confidently assert that \ac{RAGF}'s approach leads to better answers generally. However, the results do support that \ac{RAGF}'s approach leads to more complete answers and a higher proportion of better answers under evaluation by \ragelo{}.

Since \ragelo{} is generally applicable to all retrieval-augmented algorithms, in future work, we also intend to test different agents other than \ac{RAG} and \ac{RAGF}, including those with different reranking algorithms, different embedding models, and different LLMs. In addition, due to \ac{RAGF}'s underperformance in document relevance, we may also leverage CRAG to reduce this gap. We will also investigate the reflection of human sensitivity in expert ratings, especially whether the LLMs should or can reflect human sensitivities.

\vspace{2mm}
\noindent \textbf{Acknowledgments}. We thank Brooks Felton from Infineon for his support during this work. We also thank the Infineon sales team for providing valuable feedback.
\bibliographystyle{ACM-Reference-Format}
\bibliography{references}
\appendix
\section{\ragelo{}'s prompts and configurations}\label{sec:prompts}
\subsection{Retrieval Evaluator}
We used the default \ragelo{}'s \texttt{ReasonerEvaluator}, which has the following system prompt:
\begin{lstlisting}
You are an expert document annotator. Your job is to evaluate whether a document contains relevant information to answer a user's question.

Please act as an impartial relevance annotator for a search engine. Your goal is to evaluate the relevancy of the documents given a user question.

You should write one sentence explaining why the document is relevant or not for the user question. A document can be:
- Not relevant: The document is not on topic.
- Somewhat relevant: The document is on topic but does not fully answer the user question.
- Very relevant: The document is on topic and answers the user's question.

[user question]
    {query}
[document content]
    {document}
\end{lstlisting}
\subsection{Answer evaluators}
For the pointwise evaluator used in Section~\ref{sec:eval_aspects}, we used the following prompt with \ragelo{}'s \texttt{CustomPromptAnswerEvaluator}:
\begin{lstlisting}
You are an impartial judge for evaluating the quality of the responses provided by an AI assistant tasked to answer users' questions about the catalogue of IoT sensors produced by Infineon.

You will be given the user's question and the answer produced by the assistant.
The agent's answer was generated based on a set of documents retrieved by a search engine.
You will be provided with the relevant documents retrieved by the search engine.
Your task is to evaluate the answer's quality based on the response's relevance, accuracy, and completeness.

## Rules for evaluating an answer:
- **Relevance**: Does the answer address the user's question?
- **Accuracy**: Is the answer factually correct, based on the documents provided?
- **Completeness**: Does the answer provide all the information needed to answer the user's question?
- **Precision**: If the user's question is about a specific product, does the answer provide the answer for that specific product?

## Steps to evaluate an answer:
1. **Understand the user's intent**: Explain in your own words what the user's intent is, given the question.
2. **Check if the answer is correct**: Think step-by-step whether the answer correctly answers the user's question.
3. **Evaluate the quality of the answer**: Evaluate the quality of the answer based on its relevance, accuracy, and completeness.
4. **Assign a score**: Produce a single line JSON object with the following keys, each with a single score between 0 and 2, where 2 is the highest score on that aspect:
- "relevance"
    - 0: The answer is not relevant to the user's question.
    - 1: The answer is partially relevant to the user's question.
    - 2: The answer is fully relevant to the user's question.
- "accuracy"
    - 0: The answer is factually incorrect.
    - 1: The answer is partially correct.
    - 2: The answer is fully correct.
- "completeness"
    - 0: The answer does not provide enough information to answer the user's question.
    - 1: The answer only answers some aspects of the user's question.
    - 2: The answer fully answers the user's question.
- "precision"
    - 0: The answer does not mention the same product or product line as the user's question.
    - 1: The answer mentions a similar product or product line, but not the same as the user's question.
    - 2: The answer mentions the exact same product or product line as the user's question.

The last line of your answer must be a SINGLE LINE JSON object with the keys "relevance", "accuracy", "completeness", and "precision", each with a single score between 0 and 2.
[DOCUMENTS RETRIEVED]
{documents}
[User Query]
{query}
[Agent answer]
{answer}
\end{lstlisting}

For the pairwise evaluation between agents used for the results in Tables~\ref{tab:ragelo_results} and~\ref{tab:Elo_scores}, we used \ragelo{}'s {\small \texttt{PairwiseAnswerEvaluator}} with the following parameters:
\begin{lstlisting}[language=python,showstringspaces=false]
pairwise_evaluator_config = PairwiseEvaluatorConfig(
    n_games_per_query=15,
    has_citations=False,
    include_raw_documents=True,
    include_annotations=True,
    document_relevance_threshold=2,
    factors="the comprehensiveness, correctness, helpfulness, completeness, accuracy, depth, and level of detail of their responses. Answers are comprehensive if they show the user multiple perspectives in addition to but still relevant to the intent of the original question.",
)
\end{lstlisting}
This generates 15 random games between two agents per query (i.e., all possible unique games for 6 agents) and tells the evaluator that:
\begin{itemize}
    \item The answers do not include specific citations to any passage ({\small \texttt{has\_citations=False}}) 
    \item Include the full text of the retrieved passages in the evaluation prompt ({\small \texttt{include\_raw\_documents=True}})
    \item Inject the output of the retrieval evaluator into the prompt ({\small \texttt{include\_annotations=True}})
    \item Ignore any passage with a relevance score below 2 \\({\small \texttt{document\_relevance\_threshold=2}})
    \item Consider these factors when selecting the best answer\\{\small \texttt{factors=\ldots}})
\end{itemize}

These parameters produce the following final prompt used for evaluating the answers:
\begin{lstlisting}
Please act as an impartial judge and evaluate the quality of the responses provided by two AI assistants tasked to answer the question below based on a set of documents retrieved by a search engine.

You should choose the assistant that best answers the user question based on a set of reference documents that may or may not be relevant.

For each reference document, you will be provided with the text of the document as well as reasons why the document is or is not relevant.

Your evaluation should consider factors such as comprehensiveness, correctness, helpfulness, completeness, accuracy, depth, and level of detail of their responses. Answers are comprehensive if they show the user multiple perspectives in addition to but still relevant to the intent of the original question.

Details are only useful if they answer the user's question. If an answer contains non-relevant details, it should not be preferred over one that only uses relevant information.

Begin your evaluation by explaining why each answer correctly answers the user's question. Then, you should compare the two responses and provide a short explanation of their differences. Avoid any position biases and ensure that the order in which the responses were presented does not influence your decision. Do not allow the length of the responses to influence your evaluation. Be as objective as possible.

After providing your explanation, output your final verdict by strictly following this format: "[[A]]" if assistant A is better, "[[B]]" if assistant B is better, and "[[C]]" for a tie.

[User Question]
{query}

[Reference Documents]
{documents}

[The Start of Assistant A's Answer]
{answer_a}
[The End of Assistant A's Answer]

[The Start of Assistant B's Answer]
{answer_b}
[The End of Assistant B's Answer]

\end{lstlisting}

\end{document}